\input{aipcheck}

\documentclass[
    ,final            
  ]
  {aipproc}

\layoutstyle{6x9}

\begin{document}

\title{Stellar Atmospheric Parameters: The Four-Step Program and Gaia's Radial Velocity Spectrometer}


\author{Carlos Allende Prieto}{
  address={Mullard Space Science Laboratory, University College London, Holmbury St. Mary,
	Surrey RH5 6NT, United Kingdom}
}

\begin{abstract}

The determination of atmospheric parameters 
is the first and most fundamental step in the analysis of a stellar spectrum.
Current and forthcoming surveys involve samples of up to several million
stars, and therefore fully automated approaches are required to 
handle not just data reduction but also the analysis, and in particular
the determination of atmospheric parameters. We propose that a successful 
methodology needs, at the very least,  
to pass a series of consistency tests that we dub 
the 'four-step program'. This and related issues are discussed 
in some detail in the context of the massive 
data set to be obtained with the Radial Velocity Spectrometer onboard Gaia.

\end{abstract}

\maketitle


\section{Introduction}

Modern spectroscopic surveys reach data acquisition rates far superior
to those possible with conventional astronomical facilities. 
Fast speed is achieved by  
multiplexing on dedicated instruments, which demands sophisticated
data reduction pipelines. The most basic and general parts of the
data analysis must also be streamlined for a rich science return. 
In the case of stellar spectroscopy, such basic analysis covers the
determination of atmospheric parameters: surface temperature and gravity, 
and overall metallicity. Secondary parameters that may be involved are
the projected rotational velocity and the atmospheric micro- and 
macro-turbulence velocities. The determination of detailed 
chemical compositions is also amenable to automation. 

The development of automated methods for deriving atmospheric parameters 
for all types of stars is a major undertaking. The set of parameters that
can be recovered is highly dependent on the class of stars under consideration,
and the characteristics of the data.
Therefore, 
a modular approach is necessary, 
with different branches devoted to different classes of stars,
aimed at extracting the most relevant parameters in each case.


Regardless of the target classes of stars and the type of spectra and
supplementary data available, it is possible to list a number of general 
requirements that must be always met. A suitable analysis protocol must be able to 
extract the sought-after information in the spectra, and do so 
in a {\it reasonable} time. The derived quantities need to be free of
significant systematic errors, i.e. a calibration to track such systematics
must be in place. The analysis must also provide realistic  error bars, 
as expected from the uncertainties in the input data, which combined with the
systematic errors will give an adequate description of the uncertainty
in the final results. 

These requirements can be translated into a series of tests involving
both simulated data and real spectra.
In this contribution, I propose a recipe for such a validation scheme (the 
four-step program) and provide examples relevant to the European
mission Gaia, and in particular to the analysis of spectra of AFGK stars from 
the Gaia Radial Velocity Spectrometer (RVS).

\section{The Four-Step Program}

The four-step program sketched below attempts to address the most 
fundamental requirements that should be asked from an automated spectral analysis
tool. A successful protocol needs to answer affirmatively to the following questions:

\begin{enumerate}
\item Information content: Is the information we seek contained in the data?
\item Robustness: Can one recover the information with meaningful uncertainties
		in the presence of realistic noise?
\item Accuracy: Do our models resemble nature? 
\item Implementation practicalities: speed, memory, etc.: 
	Are these constraints satisfied? Are error estimates adequate?
\end{enumerate}

The design of an analysis tool will also involve an additional step, previous to
those above: the selection of the target stellar type(s), 
the relevant parameters for them,  and which parts of the available data 
set will be used. The design process will require some form of iteration
in which after part or all of the suggested checks are performed, 
 the target parameters and data selection are modified, and the tests performed again.

Most of the discussion below is generally applicable to any method
for parameter determination. It is assumed that models  are used to compute
spectra for different parameters, and one can use them to assign
 most probable values for the parameters to any given observed spectrum. 
There is some times a distinction in the literature between analyses based
on models or {\it empirical} data, but such a separation is usually artificial: 
the parameters assigned to a library 
of templates used for {\it training} or calibration 
will be ultimately linked to model atmospheres employed in earlier analyses.

With the purpose of illustrating our discussion, 
we will be using simulated data that resemble those to be obtained
by the Gaia RVS. The basic characteristics are 
a FWHM resolving power $R\equiv \lambda/\delta\lambda \simeq 11,500$
for bright targets and roughly half that value for $V>10$, 
a spectral coverage between 847 and 874 nm, and a signal-to-noise
ratio per pixel that will range from a few hundred at $V \sim 6$ to 6
at $V \sim 14$. The RVS spectral window is dominated by the Ca II
infrared triplet lines, and therefore Ca/H, more even so than Fe/H,
is a key parameter.

For our examples, we will deal with the 4-dimensional problem of
finding surface temperatures, gravities, iron and calcium abundances
from RVS spectra. We will search for the optimal solutions using a
minimum distance method based on the 
Nelder-Mead algorithm (Nelder \& Mead 1965)
and interpolation on a grid of synthetic spectra covering the spectral types A-K 
computed with Kurucz model atmospheres (Kurucz 1997).  
The search algorithm has been coded
in FORTRAN90 and described in previous reports (Allende Prieto 2004;  
Allende Prieto et al. 2006, 2008; Kilic et al. 2007).

We will consider two cases, using either absolute or normalized fluxes.
The first one corresponds to the situation when the angular diameter of a star
is known -- model atmospheres predict the flux at the stellar surface, while
observations measure it at the Earth. Alternatively, one can also use
absolute fluxes when stellar radii are adopted from stellar evolution
models and the parallaxes are known. 
For a given chemical composition, the parameters that define 
a model atmosphere (surface gravity and  effective temperature) 
can be mapped into a stellar mass and radius. We will also test the
analysis of RVS spectra normalized to the continuum.

\section{Information content}
\label{information}

The most fundamental requirement we should ask from a  protocol 
for automated determination of
stellar atmospheric parameters, is that the input data contain enough 
information to constrain the sought-after parameters. The two obstacles
to overcome are called (lack of) sensitivity and degeneracy. 
If the data have very low or zero sensitivity to the parameters of interest,
the design needs to be reconsidered. The same is true if the response of the
spectra to changes in different parameters are very similar.

One can directly test for the presence of these two pathologies using 
model spectra. The analysis protocol should return consistent and
unique solutions for the very same models used for training/calibration.
For some algorithms, e.g. neural networks, a successful 
 training process will already address this matter.


To illustrate this phase we use the aforementioned library of Gaia RVS 
synthetic spectra. We feed the same spectra to the optimization
code and compare the results with the true values used in the calculations.
The model fluxes at the edges of the grid are excluded, given that 
 the algorithm has special difficulties in those areas; there are 1875 spectra.
The mean difference (and the root-mean-squared scatter) between 
output and input iron and calcium abundances, temperatures, and gravities were,
respectively 0.0005 dex, 0.13 K, and 0.001 dex. Very similar results are 
obtained for the analysis of normalized spectra. It is interesting to
note that the widths of the 
distributions for the Fe and Ca abundances, as well as for the surface gravities, 
are dominated by truncation errors. 
In spite of this issue, the output matches 
the input parameters so tightly that we can consider that our analysis design
has succeeded in this test.  

\section{Robustness}
\label{robustness}

Real spectra will be subject to noise of multiple origins. We will divide
the sources of noise into {\it systematic} and {\it random}. It is usually
the case that there are several random components, 
following different distributions. 

The systematic errors can also include multiple
components or sources. 
For example, some manifest themselves as time-independent 
deviations between observations and models. Another kind of systematic errors,
those that can offset as a whole a model spectrum from an observed one,  
can be considered as random variables. This is the case, for example, when an angular
diameter is used to scale model fluxes computed at the stellar surface to 
observations: a random uncertainty in the angular diameter causes a
 systematic offset of the entire spectrum.

To evaluate robustness, one can use the same model spectra 
produced for the analysis, add random noise, and perform a new analysis. 
However,  
one needs to go one step further, producing new models
with different parameters. This is particularly critical for testing
the performance of different interpolation schemes, as the
results are fairly immune to interpolation errors when searching for
 solutions located at the grid nodes.


\begin{figure}
  \includegraphics[angle=90,height=.45\textheight]{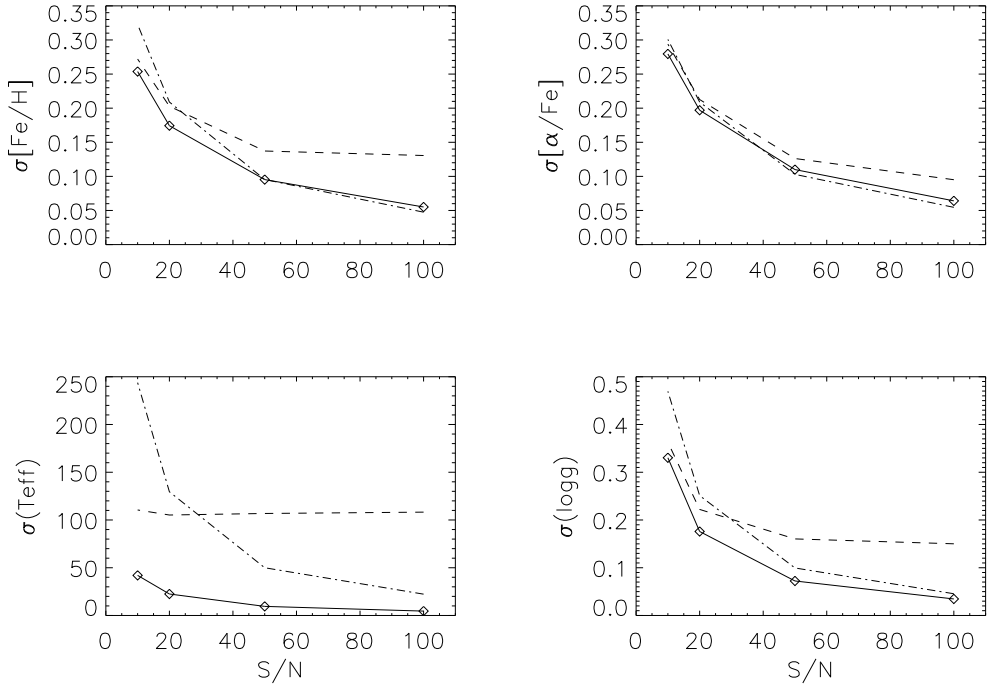}
  \caption{1$\sigma$ uncertainties in the derived parameters after adding noise
	to the model spectra. The solid lines correspond to Gaussian noise added at
	the pixel level. The dashed lines correspond to the same noise at the pixel
	level and an additional noise  
	that systematically shifts any given spectrum by a random factor from 
	a Gaussian distribution with a $\sigma=1/20$ ($S/N_{\rm sys}=20$).
	The dash-dotted lines correspond to the case of continuum-corrected spectra. 
  \label{f2}}
\end{figure}

To exemplify this phase, we go back to our Gaia RVS models and introduce
random Gaussian noise. Firstly, the spectra are degraded to a signal-to-noise
ratio per pixel of 100, 50, 20 and 10, . The results are  
summarized in 
Fig. \ref{f2}. We also consider the case when
there are additional systematic offsets that affect each spectrum's flux level
(we arbitrarily chose 
$S/N_{\rm sys}=20$), and this corresponds to the dashed lines in Fig. \ref{f2}. 
The absolute flux level is key to constrain the effective
temperatures, and as such the uncertainties for the cases when a systematic 
offset is present are much larger and independent of the signal-to-noise ratio.
On the other hand, the continuum shape has little information about abundances
and surface gravities, which makes them fairly insensitive to systematic errors.

Finally, we analyze the case in which the spectra are normalized to the continuum.
The results are also included in Fig. \ref{f2} using dash-dotted lines. Quantities 
that have a very limited effect on the continuum shape exhibit very similar error
bars as for the previous two cases, but surface temperatures are more 
poorly recovered and their uncertainties increase faster as the signal-to-noise 
degrades.

\section{Accuracy}
\label{accuracy}

Even if our system works {\it in theory}, as demonstrated after it has 
successfully  passed the tests described in the previous sections, 
subtle differences between the model spectra
and the real stellar spectra can ruin our prospects of performing 
a serious scientific analysis. 

The ultimate challenge consists in using a set of well-studied objects
with trustworthy parameters. A number of libraries 
from the literature can be used (e.g. Le Borgne et al. 2003, Valdes et al. 2004, 
Moultaka et al. 2004, Bagnulo et al. 2003,
Allende Prieto et al. 2004, S\'anchez-Bl\'azquez et al. 2006), 
but it is often the case that high quality data
are only available for few stars with a limited range of atmospheric
parameters. In this situation, a convincing demonstration requires
using several data sets. This step usually involves substantial data
processing, in order to transform existing data 
to resemble our working setup. Typical data transformations 
include adding noise and reducing the spectral resolution. 

In the case of Gaia, the use of the Near-IR spectral window restricts
the applicability of most of the existing libraries, as they tend to  
focus on  optical wavelengths. Among the obvious choices, there are 
only two possibilities: the UVES Paranal Observatory Project (Bagnulo et al. 2003) 
and the S$^4$N survey (Allende Prieto et al. 2004). Unfortunately, the
former library does not include a catalog of atmospheric parameters
and abundances. To make use of it for calibration purposes it is necessary
first to undertake a detailed analysis of the library spectra. S$^4$N includes
such a catalog, but it is affected by severe incompleteness:
it only covers about 100 stars and  the
Southern targets were observed with a 
single spectral setup that left inter-order gaps 
affecting the RVS window. This leaves us with just 64 stars.

\begin{figure}
  \includegraphics[angle=90,height=.45\textheight]{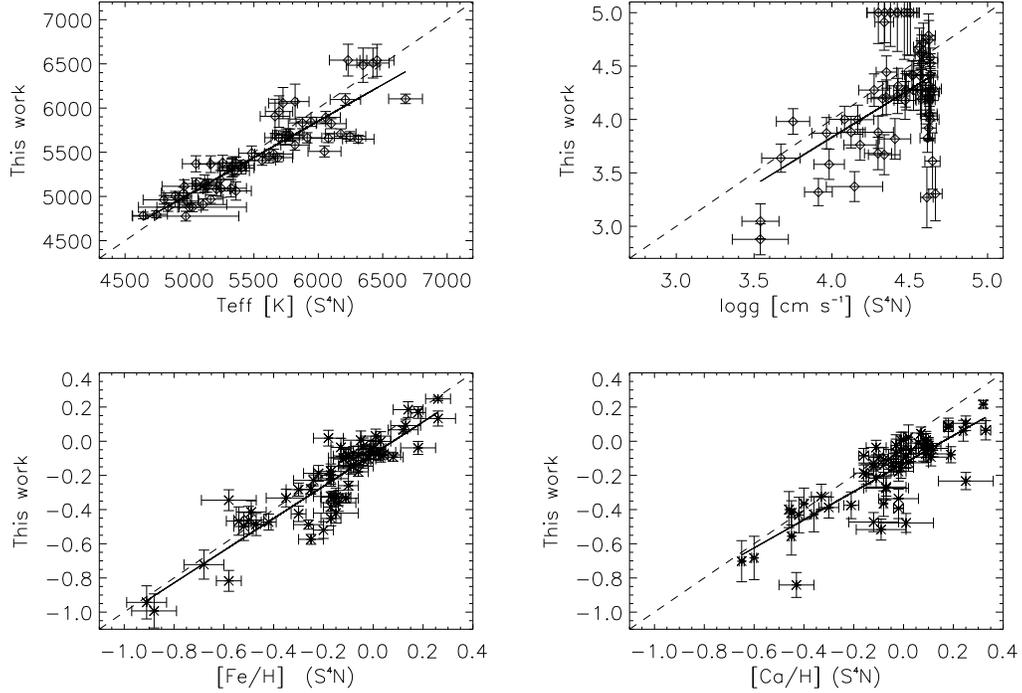}
  \caption{Comparison between the catalog  and the extracted parameters 
	for 64 stars included in the S$^4$N survey. The dashed lines are
	have a slope of one. The solid lines are least-squares fittings of
	the data to linear models.
  \label{f3}}
\end{figure}

Fig. \ref{f3} illustrates the results of the analysis of the S$^4$N  spectra.
This library does not include absolute fluxes, and therefore
we only experimented with continuum-normalized data. There is
a reasonable correlation between the catalog parameters and those
we derive from the Gaia window, but the rms 
scatter is disappointingly large:
217 K, 0.44 dex, 0.11 dex, and 0.13 dex for $T_{\rm eff}$, $\log g$, [Fe/H], and [Ca/Fe],
respectively. These figures  compare poorly 
with other results from similar or even lower resolution analysis 
(e.g. Allende Prieto et al. 2006, 2008).
Note, however, that  all these studies handle only three parameters, assuming 
a coupling between $\alpha$ and Fe.


The synthetic spectra give a fair match to the observed ones, albeit the 
agreement becomes less satisfactory for  warm F-type stars.
The limited spectral range and lines it contains does not leave 
much room for 
error. The sensitivity of the wings of the strong Ca II
triplet lines to both the calcium abundance and the stellar surface gravity 
 ties the uncertainties in these two parameters. For cool stars, where the
electron pressure is much smaller than the gas pressure and the continuum 
is dominated by the H$^-$, 
the gas pressure at a constant temperature 
is approximately proportional to 
$g^{2/3}$ (see Gray 1992). 
The wings of the Ca II lines grow proportionally 
to the calcium abundance and to $g^{1/3}$, and thus the uncertainties
in the derived surface gravities are expected to be three times larger than those
in the calcium abundance, as confirmed by the numbers above. 

The comparison between the uncertainties in the analysis of the S$^4$N spectra
and the theoretical expectations described earlier 
suggests that the synthetic spectra must become more realistic 
to allow us to cleanly disentangle the effect of the different 
atmospheric parameters, in particular gravity and
chemical abundances, from Gaia spectra of bright sources. 
Accounting for departures from LTE and/or
empirically correcting the atomic parameters to match one or several well-observed
reference spectra will likely help. The availability of absolute fluxes {\bf and}
parallaxes, will surely provide a much tighter constraint on the surface gravities
that must be taken advantage of.

\section{Implementation}
\label{implementation}

Speed and other practicalities may get in our way. In the case of Gaia,
the large number of sources places stringent constraints in the type of
 analysis that is possible within reasonable time scales. Other 
issues must also be considered, such as the required memory. 
For the Gaia tests described above, linear interpolation in the 4D 
parameter space appears accurate enough. Our tests, which involved
1122 frequencies per spectrum, used up about 50 Mbytes
of memory, mostly to hold the model grid, and took about $10^{-2}$ seconds
per spectrum on a modern processor. If suitable RVS observations
have a minimum signal-to-noise ratio per pixel of about 10, we are left
with a sample of $2-4 \times 10^7$ stars down to $V<13-14$, which could
be processed in just a few days.

Our procedure must be able to evaluate, realistically, the uncertainties 
in the estimated parameters. This can be achieved in multiple ways, but 
the most popular methods involve calculating the covariance matrix or running
 Monte Carlo simulations. Tests with both procedures
are underway for the Gaia case and will be described elsewhere.

\section{Conclusions}

In this contribution I highlight 
the need for accurate atmospheric parameters determined without human
supervision in modern stellar spectroscopic surveys. Successful programs 
must pass a series of rigorous consistency and performance checks. 
A proposal for such a set of tests is presented -- which we dub the
four-step program-- based on Monte Carlo simulations and libraries
of real observed spectra. The program will need to be customized for
each particular data set and analysis algorithm, but the basics are
likely of general application.

We illustrate the different phases of the four-step program with 
simulated data for the Gaia RVS. It is pointed
out that there is no existing spectral library of observations 
that is up to the task
of calibrating the tools for deriving atmospheric parameters 
for this particular data set. An appropriate catalog of high-resolution spectra 
should be assembled. A consistent spectral analysis of the targets included
in such library should be carried 
afresh, taking advantage of a wider spectral range (including the visible), 
in order to avoid inheriting systematic errors introduced by combining results
from multiple sources in the literature.





\bibliographystyle{aipproc}   

\bibliography{sample}





\end{document}